\documentclass[a4paper,10pt,twoside]{cpc-hepnp}
\usepackage{CJK,upgreek,fancyhdr}
\usepackage{multicol}
\usepackage{graphicx}
\usepackage{booktabs}
\usepackage{amssymb,bm,mathrsfs,bbm,amscd}
\usepackage[tbtags]{amsmath}
\usepackage{lastpage}

\begin{document}


\footnotetext[0]{Received xxxx 2016}

\title{Investigation of the high-spin rotational properties of the proton emitter
       $^{113}$Cs using a particle-number conserving method
       \thanks{Supported by National Natural Science
               Foundation of China (11275098, 11275248, 11505058) and
               Fundamental Research Funds for the Central Universities (2015QN21)}}

\author{%
      Zhen-Hua Zhang $^{1)}$\email{zhzhang@ncepu.edu.cn}%
\quad Peng Xu 
}
\maketitle

\address{Mathematics and Physics Department,
         North China Electric Power University, Beijing 102206, China
}

\begin{abstract}

The recently observed two high-spin rotational bands in the proton
emitter $^{113}$Cs are investigated using the cranked shell model
with pairing correlations treated by a particle-number conserving method,
in which the Pauli blocking effects are taken into account exactly.
By using the configuration assignments of
band 1 ($\pi 3/2^+[422], \alpha = -1/2$)
and band 2 ($\pi 1/2^+[420], \alpha = 1/2$),
the experimental moments of inertia and quasiparticle alignments
can be well reproduced by the present calculations,
which in turn strongly support these configuration assignments.
Furthermore, by analyzing the occupation probability $n_\mu$ of each cranked
Nilsson level near the Fermi surface and the contribution of
each orbital to the angular momentum alignments,
the backbending mechanism of these two bands is also investigated.

\end{abstract}




\begin{multicols}{2}

\section{Introduction}
\label{sec:intro}

Investigation of nuclei far from the $\beta$-stability line is one of the
most important frontiers in nuclear physics.
Nuclei at the extremes of stability show various exotic decay modes,
which can provide valuable information for the study
of the nuclear structure close to the drip-line.
On the proton-rich side of the nuclear landscape,
the existence of the Coulomb potential together with the centrifugal potential,
gives rise to barriers of about 15~MeV.
Therefore, relatively long-lived proton emitters,
lifetimes ranging from $10^{-6}$~s to a few seconds, can exist beyond the proton
drip-line~\cite{Blank2008_PPNP60-403, Pfutzner2012_RMP84-567}.
Experimentally, after the first direct emission of a proton from a
isomeric state was observed in $^{53}$Co~\cite{Jackson1970_PLB33-281},
up to now, almost 50 proton emitters, including
the one-proton emitters with charges in the range $50<Z<83$,
two-proton emitters below $Z=50$~\cite{Giovinazzo2002_PRL89-102501},
and $\beta$-delayed proton emitters~\cite{Borge2002_NPA701-373}, {\it etc.},
have been identified.
Theoretically, various microscopic models have been used to explain
the ground state properties, measured half-lives and spectroscopic information
of the proton emitters~\cite{Aberg1997_PRC56-1762, Lalazissis1999_NPA650-133,
Fiorin2003_PRC67-054302, Olsen2013_PRL110-222501, Ferreira2016_PLB753-237}.

Most of the observed proton emitters have a spherical shape.
Anomalous proton decay rates have been measured for
$^{109}$I and $^{113}$Cs, which are consistent with calculations
assuming relatively small deformations~\cite{Gillitzer1987_ZPA326-107}.
Furthermore, experimental efforts to investigate proton emitters in the
light rare-earth region have brought to light the existing of the deformed
nuclei at the drip-line~\cite{Davids1998_PRL80-1849, Sonzogni1999_PRL83-1116}
and the rotational bands have also been observed~\cite{Seweryniak2001_PRL86-1458}.
The lifetimes of these deformed proton emitters can provide direct information on the
last occupied Nilsson orbital and the shape of the nucleus.
Recently, two previously observed high-spin rotational bands in the deformed proton emitter
$^{113}$Cs have been extended to spins of $45/2~\hbar$ and $51/2~\hbar$, respectively~\cite{Wady2015_PLB740-243}.
The excitation energies of these two bands are over 8~MeV above the ground state~\cite{Wady2015_PLB740-243}.
Up to now, these are the highest spins and excitation energies
observed in the nuclei beyond the proton drip-line.
As one of the first proton emitters observed,
decay properties of $^{113}$Cs have been given by the most recent investigations
as half-life $T_{1/2}=16.7(7)$ ${\rm\mu s}$~\cite{Batchelder1998_PRC57-R1042}
and proton energy $E_{\rm p}=959(6)$~keV~\cite{Page1994_PRL72-1798}.
The investigation of the rotational bands observed in $^{113}$Cs allow extracting of
those properties such as moments of inertia (MOI's) and backbending frequencies,
which provide a bentchmark for various nuclear models,
{\it e.g.}, the cranked Nilsson-Strutinsky method~\cite{Andersson1976_NPA268-205},
the Hartree-Fock-Bogoliubov cranking model with Nilsson potential~\cite{Bengtsson1979_NPA327-139}
and Woods-Saxon potential~\cite{Nazarewicz1985_NPA435-397, Cwiok1987_CPC46-379},
the tilted axis cranking model~\cite{Frauendorf2001_RMP73-463},
the cranked relativistic~\cite{Afanasjev1996_NPA608-107}
and non-relativistic mean-field models~\cite{Dobaczewski1997_CPC102-166},
the projected shell model~\cite{Hara1995_IJMPE4-637}, {\it etc}.

In this paper, the cranked shell model (CSM) with
pairing correlations treated by a particle-number conserving
(PNC) method~\cite{Zeng1983_NPA405-1, Zeng1994_PRC50-1388}
is used to investigate the two high-spin rotational bands recently observed in
the proton emitter $^{113}$Cs~\cite{Wady2015_PLB740-243}.
In contrary to the conventional Bardeen-Cooper-Schrieffer or
Hartree-Fock-Bogolyubov approaches, in the PNC method, the Hamiltonian is
diagonalized directly in a truncated Fock-space~\cite{Wu1989_PRC39-666}.
So the particle-number is conserved and the Pauli blocking effects are treated exactly.
The PNC-CSM has already been used successfully for describing the odd-even differences
in MOI's~\cite{Zeng1994_PRC50-746},
the identical bands~\cite{Liu2002_PRC66-024320, He2004_CPL21-813, He2004_CPC28-1366, He2005_EPJA23-217},
the nonadditivity in MOI's~\cite{Liu2002_PRC66-067301, He2005_NPA760-263, Zhang2008_ChinPhysC32-681},
the nuclear pairing phase transition~\cite{Wu2011_PRC83-034323},
the high-spin rotational bands in the rare-earth~\cite{Liu2004_NPA735-77, Zhang2009_NPA816-19,
Zhang2009_PRC80-034313, Zhang2010_ChinPhysC34-39, Zhang2010_ChinPhysC34-1836,
Li2013_ChinPhysC37-014101, Zhang2016_NPA949-22, Zhang2016_SciChinaPMA59-672012},
the actinide and superheavy nuclei~\cite{He2009_NPA817-45, Zhang2011_PRC83-011304R,
Zhang2012_PRC85-014324, Zhang2013_PRC87-054308, Li2016_SciChinaPMA59-672011},
and the nuclear antimagnetic rotation~\cite{Zhang2013_PRC87-054314}.
Note that the PNC scheme has been implanted both in relativistic
and nonrelativistic mean field models~\cite{Meng2006_FPC1-38, Pillet2002_NPA697-141} and
the total-Routhian-surface method with the
Woods-Saxon potential~\cite{Fu2013_PRC87-044319, Fu2013_SCPMA56-1423}.
Very recently, PNC method based on the
cranking Skyrme-Hartree-Fock model has been developed~\cite{Liang2015_PRC92-064325}.

This paper is organized as follows.
A brief introduction to the PNC treatment of pairing correlations within
the CSM is presented in Sec.~2.
The numerical details used in the PNC-CSM calculation are given in Sec.~3.
This method is used to investigate the two rotational bands of $^{113}$Cs in Sec.~4.
A brief summary is given in Sec.~5.

\section{A brief introduction to particle-number conserving method for the cranked shell model}
\label{sec:pnc}

The cranked shell model Hamiltonian of an axially symmetric
nucleus in the rotating frame can be written as
\begin{eqnarray}
 H_\mathrm{CSM}
 & = &
 H_0 + H_\mathrm{P}
 = H_{\rm Nil}-\omega J_x + H_\mathrm{P}
 \ ,
 \label{eq:H_CSM}
\end{eqnarray}
where $H_{\rm Nil}$ is the Nilsson Hamiltonian, $-\omega J_x$ is the
Coriolis interaction with the cranking frequency $\omega$ about the
$x$ axis (perpendicular to the nuclear symmetry $z$ axis).
$H_{\rm P}$ is the pairing interaction,
\begin{eqnarray}
 H_{\rm P}
 & = &
  -G \sum_{\xi\eta} a^\dag_{\xi} a^\dag_{\bar{\xi}}
                        a_{\bar{\eta}} a_{\eta}
  \ ,
\end{eqnarray}
where $\bar{\xi}$ ($\bar{\eta}$) labels the time-reversed state of a
Nilsson state $\xi$ ($\eta$),
and $G$ is the effective strength of monopole pairing interaction.

Instead of the usual single-particle level truncation in conventional
shell-model calculations, a cranked many-particle configuration (CMPC)
truncation (Fock space truncation) is adopted which is crucial
to make the particle-number conserving calculations for low-lying excited states both
workable and sufficiently accurate~\cite{Molique1997_PRC56-1795, Wu1989_PRC39-666}.
Usually a dimension of 1000 should be enough for the calculations of heavy nuclei.
An eigenstate of $H_\mathrm{CSM}$ can be written as
\begin{equation}
 |\Psi\rangle = \sum_{i} C_i \left| i \right\rangle
 \qquad (C_i \; \textrm{real}) \ ,
\end{equation}
where $| i \rangle$ is a CMPC (an eigenstate of the one-body operator $H_0$).
By diagonalizing the $H_\mathrm{CSM}$ in a sufficiently
large CMPC space, sufficiently accurate solutions for low-lying excited eigenstates of
$H_\mathrm{CSM}$ are obtained.

The angular momentum alignment for the state $| \Psi \rangle$ is
\begin{equation}
\langle \Psi | J_x | \Psi \rangle = \sum_i C_i^2 \langle i | J_x | i
\rangle + 2\sum_{i<j}C_i C_j \langle i | J_x | j \rangle \ ,
\end{equation}
and the kinematic MOI of state $| \psi \rangle$ is
\begin{equation}
J^{(1)}=\frac{1}{\omega} \langle\Psi | J_x | \Psi \rangle \ .
\end{equation}
Because $J_x$ is a one-body operator, the matrix element $\langle i | J_x | j \rangle$
($i\neq j$) may not vanish only when
$|i\rangle$ and $|j\rangle$ differ by
one particle occupation~\cite{Zeng1994_PRC50-1388}.
After a certain permutation of creation operators,
$|i\rangle$ and $|j\rangle$ can be recast into
\begin{equation}
 |i\rangle=(-1)^{M_{i\mu}}|\mu\cdots \rangle \ , \qquad
|j\rangle=(-1)^{M_{j\nu}}|\nu\cdots \rangle \ ,
\end{equation}
where $\mu$ and $\nu$ denotes two different single-particle states,
and $(-1)^{M_{i\mu}}=\pm1$, $(-1)^{M_{j\nu}}=\pm1$ according to
whether the permutation is even or odd.
Therefore, the angular momentum alignment of
$|\Psi\rangle$ can be expressed as
\begin{equation}
 \langle \Psi | J_x | \Psi \rangle = \sum_{\mu} j_x(\mu) + \sum_{\mu<\nu} j_x(\mu\nu)
 \ .
 \label{eq:jx}
\end{equation}
where the diagonal contribution $j_x(\mu)$ and the
off-diagonal (interference) contribution $j_x(\mu\nu)$ can be written as
\begin{eqnarray}
j_x(\mu)&=&\langle\mu|j_{x}|\mu\rangle n_{\mu} \ ,
\\
j_x(\mu\nu)&=&2\langle\mu|j_{x}|\nu\rangle\sum_{i<j}(-1)^{M_{i\mu}+M_{j\nu}}C_{i}C_{j}
  \quad  (\mu\neq\nu) \ ,
\end{eqnarray}
and
\begin{equation}
n_{\mu}=\sum_{i}|C_{i}|^{2}P_{i\mu} \ ,
\end{equation}
is the occupation probability of the cranked orbital $|\mu\rangle$,
$P_{i\mu}=1$ if $|\mu\rangle$ is occupied in $|i\rangle$, and
$P_{i\mu}=0$ otherwise.

\section{Numerical details}
\label{sec:num}

In this work, the Nilsson parameters ($\kappa$ and $\mu$) for $^{113}$Cs are taken
from the traditional values~\cite{Bengtsson1985_NPA436-14}.
The deformation parameters $\varepsilon_2 = 0.192$ and $\varepsilon_4 = -0.027$
are taken from Ref.~\cite{Moeller1995_ADNDT59-185}.
The valence single-particle space in this work is constructed
in the major shells from $N=3$ to $N=5$ both for protons and neutrons.
In principle, the effective pairing strengths can
be determined by the odd-even differences in nuclear binding energies,
and are connected with the dimension of the truncated CMPC space.
The CMPC truncation energies are
about 0.9$\hbar\omega_0$ both for protons and neutrons.
For $^{113}$Cs, $\hbar\omega_{\rm 0p}=8.406$~MeV
for protons and $\hbar\omega_{\rm 0n}=8.556$~MeV for neutrons~\cite{Nilsson1969_NPA131-1}.
The dimensions of the CMPC space are about 1000 both for protons and neutrons.
The corresponding effective monopole pairing strengths used in this work are
$G_{\rm p}$ = 0.5~MeV and $G_{\rm n}$ = 0.7~MeV.
A larger CMPC space with renormalized effective pairing strengths gives essentially the same results.
In addition, the stability of the PNC-CSM calculations against the change
of the dimension of the CMPC space has been investigated in
Refs.~\cite{Zeng1994_PRC50-1388, Zhang2012_PRC85-014324}.
In present calculations, almost all the important CMPC's
(with the corresponding weights larger than $0.1\%$) are taken into account,
so the solutions to the low-lying excited states are accurate enough.

\section{Results and discussion}
\label{sec:resu}

\end{multicols}
\begin{center}
\includegraphics[width=0.9\columnwidth]{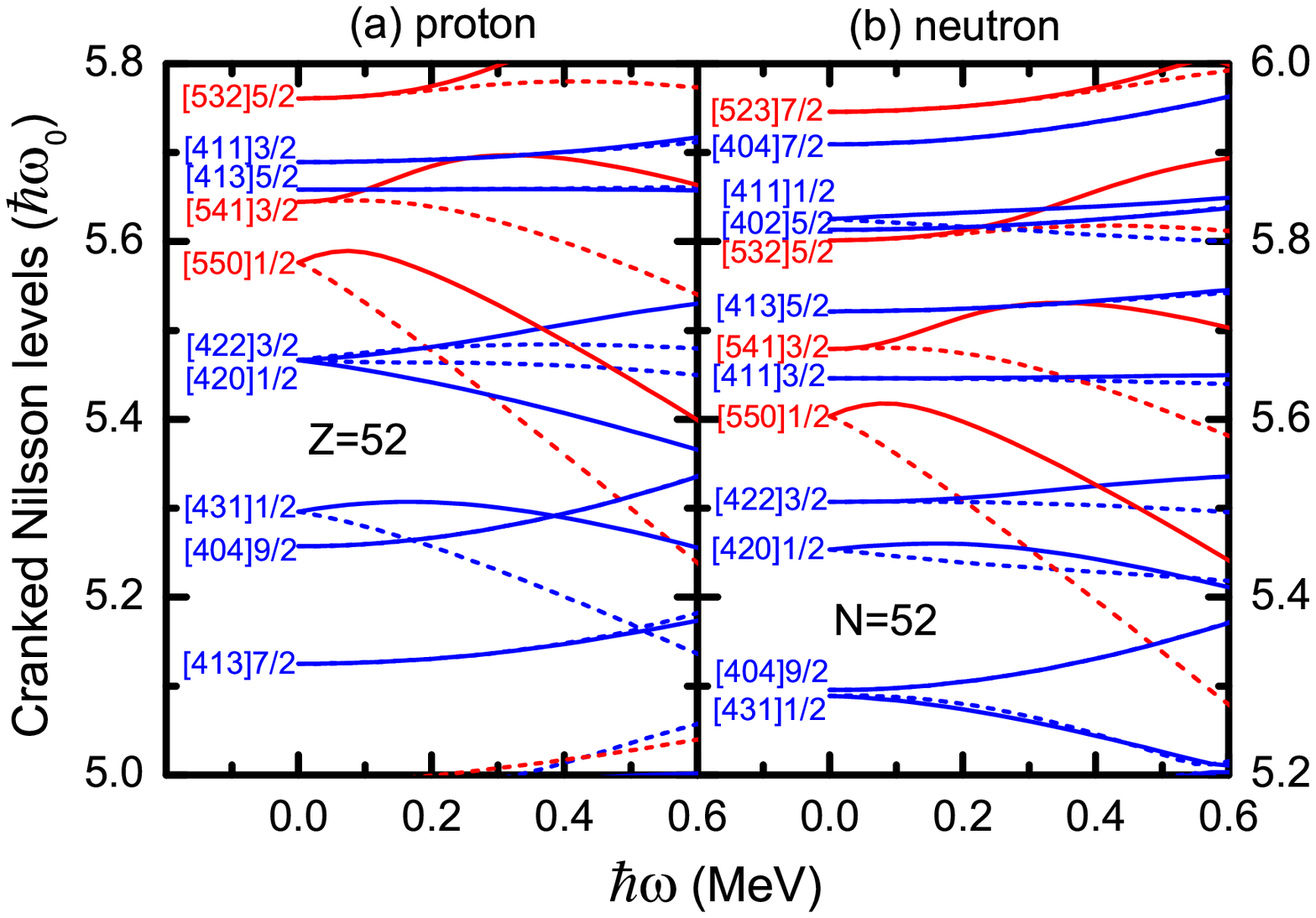}
\figcaption{\label{fig1:Nil}(Color online)
The cranked Nilsson levels near the Fermi surface of $^{113}$Cs
for (a) protons and (b) neutrons.
The positive (negative) parity levels are denoted by blue (red) lines.
The signature $\alpha=+1/2$ ($\alpha=-1/2$) levels
are denoted by solid (dotted) lines.
The Nilsson parameters ($\kappa$ and $\mu$) are taken from
the traditional values~\cite{Bengtsson1985_NPA436-14}.
The deformation parameters $\varepsilon_2 = 0.192$ and $\varepsilon_4 = -0.027$
are taken from Ref.~\cite{Moeller1995_ADNDT59-185}.}
\end{center}
\begin{multicols}{2}

Figure~\ref{fig1:Nil} shows the calculated cranked Nilsson levels near the
Fermi surface of $^{113}$Cs for protons and neutrons.
The positive (negative) parity levels are denoted by blue (red) lines.
The signature $\alpha=+1/2$ ($\alpha=-1/2$) levels
are denoted by solid (dotted) lines.
It can be seen from Fig.~\ref{fig1:Nil}(a) that in the present calculation,
the ground state of $^{113}$Cs is $\pi 3/2^+[422]$ ($2d_{5/2}$), which is consistent with the
ground state assignment $3/2^+$ in Refs.~\cite{Delion2006_PRL96-072501, Wady2015_PLB740-243},
and the lowest-lying negative parity state is $\pi 1/2^-[550]$ ($h_{11/2}$).
The energy of the first excited state $\pi 1/2^+[420]$ ($g_{7/2}$),
which is the pseudospin partner of $\pi 3/2^+[422]$,
is very close to that of the ground state.
The situation is similar for the higher excited states
$\pi 3/2^+[411]$ ($g_{7/2}$) and $\pi 5/2^+[413]$ ($2d_{5/2}$).
The near degeneracy of these pseudospin partners may indicate the existence of the pseudospin symmetry
in this axially well deformed proton emitter nucleus,
since it has been pointed out in Ref.~\cite{Liang2015_PR570-1} that
a better pseudospin symmetry can be expected for nuclei close to the proton and neutron drip-lines.
It should be noted that in the present calculation, the orbitals $\pi 1/2^+[420]$ and
$\pi 3/2^+[422]$ are closer to the Fermi surface than $\pi 1/2^-[550]$,
which is consistent with the Woods-Saxon CSM results in Ref.~\cite{Wady2015_PLB740-243}.
In the following, this cranked Nilsson level scheme will be adopted to
investigate the rotational bands recently observed in the proton emitter nucleus $^{113}$Cs.

\begin{center}
\includegraphics[width=0.99\columnwidth]{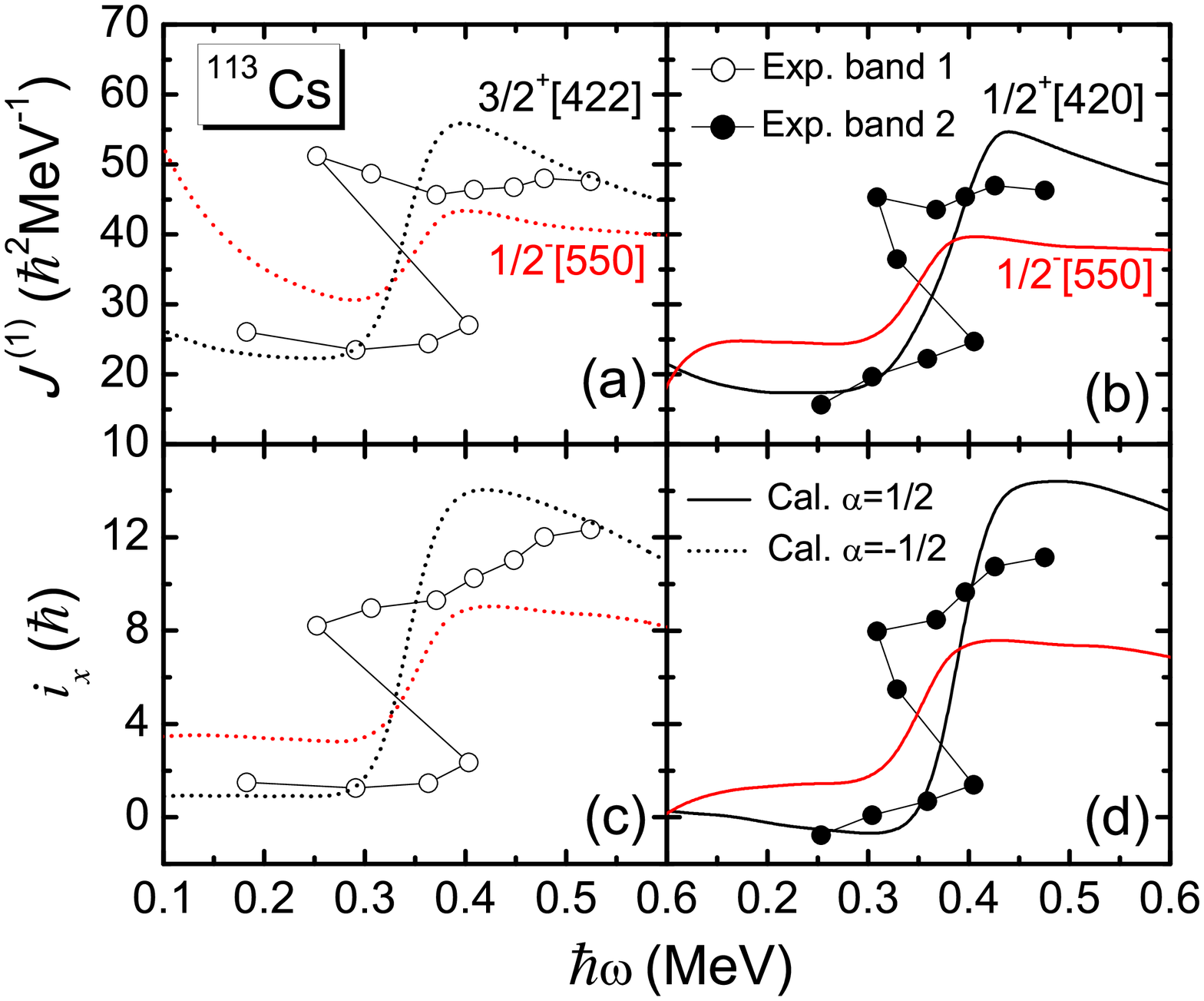}
\figcaption{\label{fig2:MOI}(Color online)
The experimental and calculated kinematic MOI's
$J^{(1)}$ and alignments of band 1 ($\pi 3/2^+[422], \alpha = -1/2$)
and band 2 ($\pi 1/2^+[420], \alpha = 1/2$) in $^{113}$Cs.
The data are taken from Ref.~\cite{Wady2015_PLB740-243}.
The alignments $i$ are defined as $i= \langle
J_x \rangle -\omega J_0 -\omega ^ 3 J_1$ and the Harris parameters
$J_0 = 17.0\ \hbar^2$MeV$^{-1}$ and $J_1 = 25.8\ \hbar^4$MeV$^{-3}$ are
taken from Ref.~\protect\cite{Wady2015_PLB740-243}.
The experimental MOI's and alignments are denoted by solid circles (signature
$\alpha=+1/2)$ and open circles (signature $\alpha=-1/2)$, respectively.
The calculated MOI's and alignments are
denoted by black solid lines (signature $\alpha=+1/2)$ and black dotted lines
(signature $\alpha=-1/2)$, respectively.
The calculated results with $\pi 1/2^-[550]$, $\alpha=\pm 1/2$ are also shown as red lines.
}
\end{center}

Figure~\ref{fig2:MOI} shows the experimental and calculated kinematic MOI's
$J^{(1)}$ and alignments $i$ of band 1 ($\pi 3/2^+[422], \alpha = -1/2$)
and band 2 ($\pi 1/2^+[420], \alpha = 1/2$) in $^{113}$Cs.
The configuration assignments and the data are taken from Ref.~\cite{Wady2015_PLB740-243}.
The alignments $i$ are defined as $i= \langle
J_x \rangle -\omega J_0 -\omega ^ 3 J_1$ and the Harris parameters
$J_0 = 17.0\ \hbar^2$MeV$^{-1}$ and $J_1 = 25.8\ \hbar^4$MeV$^{-3}$ are
taken from Ref.~\cite{Wady2015_PLB740-243}.
The experimental MOI's and alignments are denoted by solid circles (signature
$\alpha=+1/2)$ and open circles (signature $\alpha=-1/2)$, respectively.
The calculated MOI's and alignments are
denoted by black solid lines (signature $\alpha=+1/2)$ and black dotted lines
(signature $\alpha=-1/2)$, respectively.
In previous investigations~\cite{Gross1998_AIP455-444, Yu2003_AIP681-172},
the rotational bands observed in $^{113}$Cs are assigned as $\pi h_{11/2}$.
To make clear the configuration assignments for these two rotational bands,
the calculated results with $\pi 1/2^-[550]$ ($h_{11/2}$), $\alpha=\pm 1/2$,
are also shown as red lines for comparison.
It can be seen from Fig.~\ref{fig2:MOI} that the experimental MOI's and alignments
of these two rotational bands and their variation
with rotational frequency $\hbar\omega$ are well reproduced by the
PNC calculations using the configuration assignments in Ref.~\cite{Wady2015_PLB740-243}
except a little lager results than the data after backbending,
while the calculated results using the configuration $\pi 1/2^-[550]$ deviate a lot from the data.
Therefore, the present calculations support the configuration assignments
of band 1 ($\pi 3/2^+[422], \alpha = -1/2$) and band 2 ($\pi 1/2^+[420], \alpha = 1/2$)
in Ref.~\cite{Wady2015_PLB740-243}.
It should be noted that for $^{113}$Cs with neutron number $N=Z+3$,
the neutron-proton pairing correlations may play an important role
on the properties of rotational alignments in the high-$j$
proton and neutron $h_{11/2}$
subshell~\cite{Frauendorf1999_PRC59-1400, Frauendorf2014_PPNP78-24}.
After considering this effect, the calculated results may be improved.
Moreover, the sharp backbendings at $\hbar\omega \sim 0.35$~MeV
in the experimental MOI's and alignments for band 1 and band 2
are also not very well reproduced by the calculation.
This is because in the cranking model, before and after the backbending,
the two bands which have quite different quasiparticle alignment from each other are mixed.
In order to obtain the backbending effect exactly,
one has to go beyond the cranking model and consider the two quasiparticle configurations
in the vicinity of the backbending region~\cite{Hamamoto1976_NPA271-15, Cwiok1978_PLB76-263}.

It is well known that the backbending is caused by the alignment of the high-$j$
intruder orbitals~\cite{Stephens1972_NPA183-257},
which corresponding to the proton and neutron $h_{11/2}$ orbitals
in $A \sim 110$ mass region.
For band 1 ($\pi 3/2^+[422], \alpha = -1/2$)
and band 2 ($\pi 1/2^+[420], \alpha = 1/2$) in $^{113}$Cs,
the proton $\pi h_{11/2}$ orbitals are not blocked.
Therefore, both the proton and the neutron $h_{11/2}$ orbitals may contribute
to the alignment after the backbending.
One of the advantages of the PNC method is that the
total particle number $N = \sum_{\mu}n_\mu$ is exactly conserved,
whereas the occupation probability $n_\mu$ for each orbital varies
with rotational frequency $\hbar\omega$.
By examining the $\omega$-dependence of the orbitals close to the Fermi surface, one
can learn more about how the Nilsson levels evolve with rotation and
get some insights on the backbendings.
Figure~\ref{fig3:Occup} shows the occupation probability $n_\mu$ of each orbital
$\mu$ (including both $\alpha=\pm1/2$)
near the Fermi surface for the band 1 ($\pi 3/2^+[422], \alpha = -1/2$)
and band 2 ($\pi 1/2^+[420], \alpha = 1/2$) in $^{113}$Cs.
The positive and negative parity levels are denoted
by blue solid and red dotted lines, respectively.
The Nilsson levels far above the Fermi surface
($n_{\mu}\sim0$) and far below ($n_{\mu}\sim2$) are not shown.
It can be seen from Fig.~\ref{fig3:Occup}(a) that at
the rotational frequency $\hbar\omega \sim 0.35$~MeV,
occupation probabilities of the orbital $\nu 1/2^-[550]$ increase quickly from 0.8 to about 2.0,
while the occupation probabilities of some other orbitals, {\it e.g.},
$\nu 3/2^-[541]$, $\nu 3/2^+[411]$ and $\nu 3/2^+[422]$, slightly decrease.
This indicate that for band 1 and band 2, the contribution to the backbending in neutrons
mainly comes from the $\nu h_{11/2}$ orbitals.
Fig.~\ref{fig3:Occup}(b) shows that the occupation probability of
the orbital $\pi 1/2^-[550]$ increases quickly from 0.5 to about 2.0, while
the occupation probability of $\pi 1/2^+[420]$ decreases from 1.2 to about 0.4.
This indicate that for band 1, the contribution to the backbending in protons
mainly comes from the $\pi h_{11/2}$ orbitals.
It also can be found that the backbending frequencies in protons and neutrons
are very close to each other in band 1, which is consistent with
the Woods-Saxon CSM calculations in Ref.~\cite{Wady2015_PLB740-243}.
The proton occupation probability for band 2 ($\pi 1/2^+[420], \alpha = 1/2$)
in Fig.~\ref{fig3:Occup}(c) is very similar with that of band 1,
except a little latter backbending frequency.
Note that in both band 1 and band 2, the pseudospin partner orbitals
$\pi 3/2^+[411]$ and $\pi 1/2^+[420]$ are mixed before the backbending.
Therefore, the rotational properties of these two bands, {\it e.g.}, MOI's and alignments,
are very similar to each other.

\end{multicols}
\begin{center}
\includegraphics[width=0.9\columnwidth]{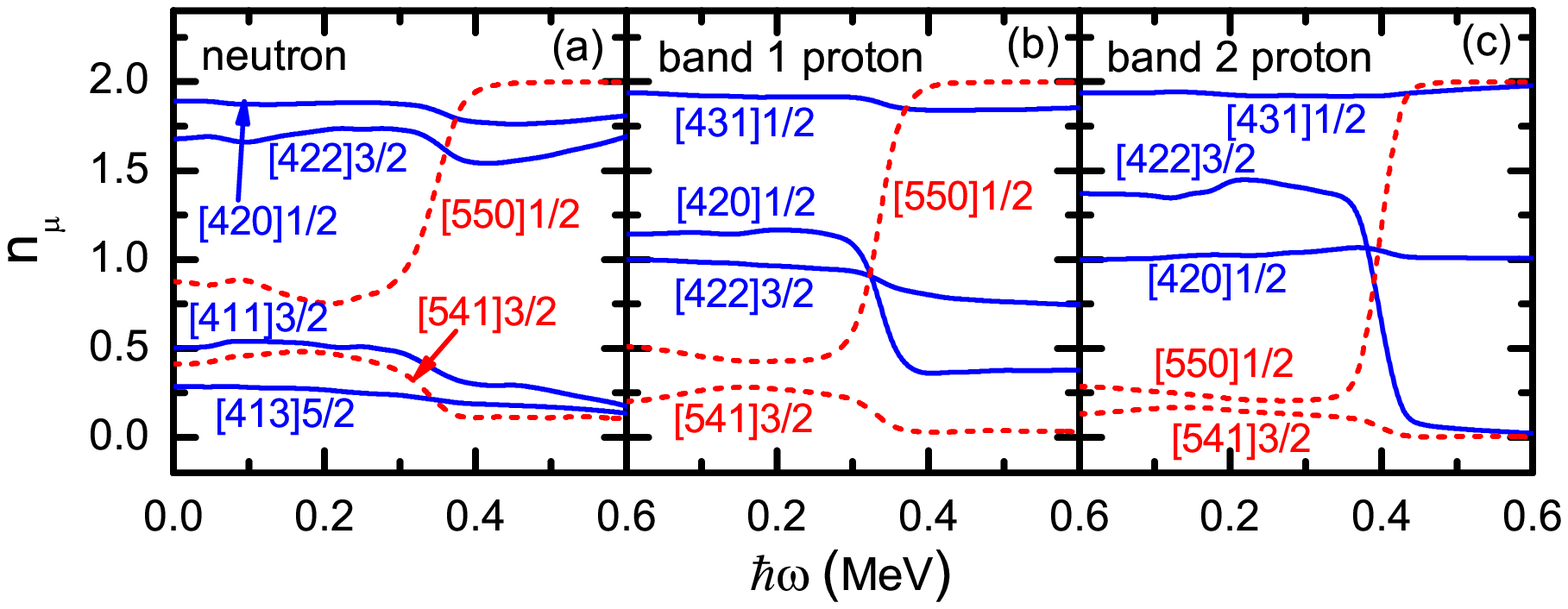}
\figcaption{\label{fig3:Occup}(Color online)
Occupation probability $n_\mu$ of each orbital $\mu$ (including both $\alpha=\pm1/2$)
near the Fermi surface for the band 1 ($\pi 3/2^+[422], \alpha = -1/2$)
and band 2 ($\pi 1/2^+[420], \alpha = 1/2$) in $^{113}$Cs.
The positive (negative) parity levels are denoted by blue solid (red dotted) lines.
The Nilsson levels far above the Fermi surface
($n_{\mu}\sim0$) and far below ($n_{\mu}\sim2$) are not shown.
}
\end{center}
\begin{multicols}{2}

\end{multicols}
\begin{center}
\includegraphics[width=0.9\columnwidth]{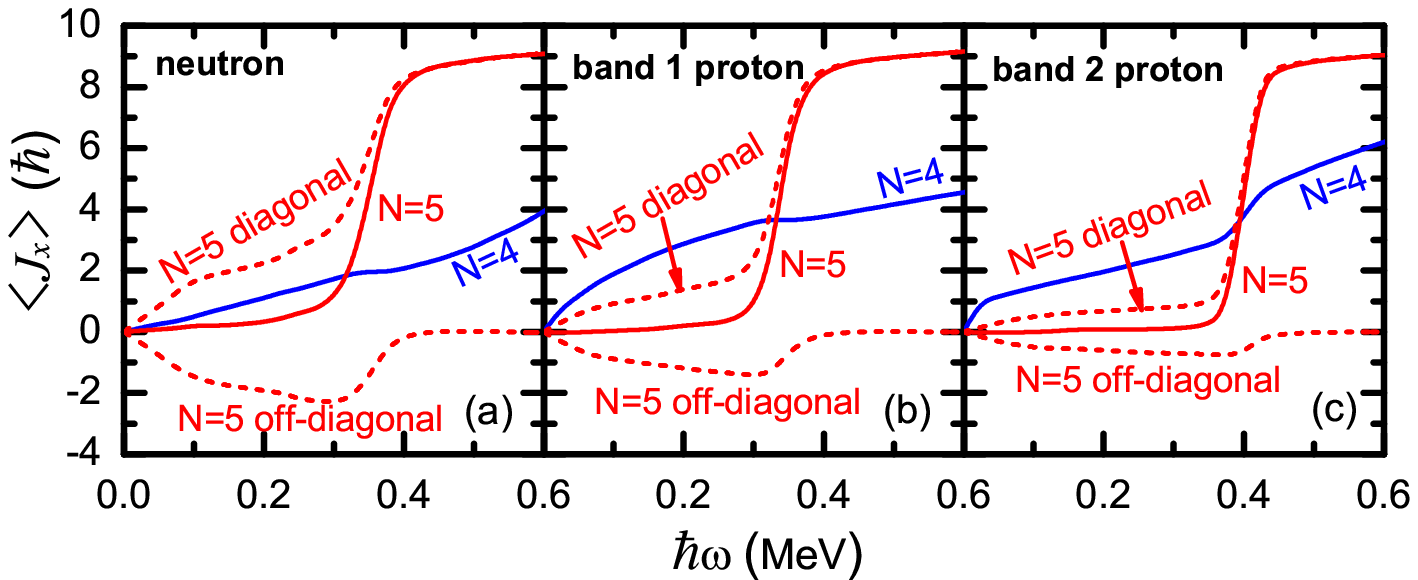}
\figcaption{\label{fig4:Jxshell}(Color online)
Contributions of each proton and neutron major shell to the angular momentum
alignment $\langle J_x\rangle$ for the band 1 ($\pi 3/2^+[422], \alpha = -1/2$)
and band 2 ($\pi 1/2^+[420], \alpha = 1/2$) in $^{113}$Cs.
The diagonal $\sum_{\mu} j_x(\mu)$ and
off-diagonal parts $\sum_{\mu<\nu} j_x(\mu\nu)$ in
Eq.~(\protect\ref{eq:jx}) from the proton $N=5$
shells are shown by dotted lines.%
}
\end{center}
\begin{multicols}{2}

In Fig.~\ref{fig4:Jxshell}, the contributions of each proton and neutron
major shell to the angular momentum alignment $\langle J_x\rangle$ for
the band 1 ($\pi 3/2^+[422], \alpha = -1/2$)
and band 2 ($\pi 1/2^+[420], \alpha = 1/2$) in $^{113}$Cs are shown.
The diagonal $\sum_{\mu} j_x(\mu)$ and
off-diagonal parts $\sum_{\mu<\nu} j_x(\mu\nu)$ in
Eq.~(\protect\ref{eq:jx}) from the proton $N=5$
shells are shown by dotted lines.
Note that in this figure, the smoothly increasing part of the
alignment represented by the Harris formula
($\omega J_0 +\omega^ 3 J_1$) is not subtracted.
It can be seen clearly that for both neutrons and protons in band 1 and 2,
the angular momentum alignments after the backbending mainly come from the
$N=5$ major shell.
Moreover, for neutrons [Fig.~\ref{fig4:Jxshell}(a)], both the diagonal and the off-diagonal parts
contribute to the backbending.
While for protons in band 1 [Fig.~\ref{fig4:Jxshell}(b)],
the diagonal part have more contribution than the off-diagonal part,
which become much smaller than the diagonal part in band 2 [Fig.~\ref{fig4:Jxshell}(c)].

In order to have a  more clear understanding of the backbending
mechanism, the contributions of each proton and neutron orbital in the $N=5$
major shell to the angular momentum alignments
$\langle J_x\rangle$ for the band 1 ($\pi 3/2^+[422], \alpha = -1/2$)
and band 2 ($\pi 1/2^+[420], \alpha = 1/2$) in $^{113}$Cs are shown in Fig.~\ref{fig5:Jxorb}.
The diagonal (off-diagonal) part $j_x(\mu)$ [$j_x(\mu\nu)$]
in Eq.~(\protect\ref{eq:jx}) is denoted by blue solid (red dotted) lines.
In Fig.~\ref{fig5:Jxorb}(a) one can easily find that for neutrons,
the diagonal part $j_x(\nu 1/2^-[550])$ and the off-diagonal parts
$j_x(\nu1/2^-[550]\nu3/2^-[541])$ and $j_x(\nu3/2^-[541]\nu5/2^-[532])$ change a lot
after the backbending ($\hbar\omega\sim$ 0.35~MeV).
The alignment gain after the upbending mainly comes from these terms.
In Fig.~\ref{fig5:Jxorb}(b) for protons in band 1, the contribution from the
diagonal part $j_x(\pi1/2^-[550])$ is much larger than the off-diagonal parts
$j_x(\pi1/2^-[550]\pi3/2^-[541])$ and $j_x(\pi3/2^-[541]\pi5/2^-[532])$.
While for the protons in band 2 [Fig.~\ref{fig5:Jxorb}(c)],
the contribution from the off-diagonal parts are negligible.
Therefore, it can be understand that even band 1 and band 2 are pseudospin partners,
the rotational properties are a little different due to the interference terms.

\end{multicols}
\begin{center}
\includegraphics[width=0.9\columnwidth]{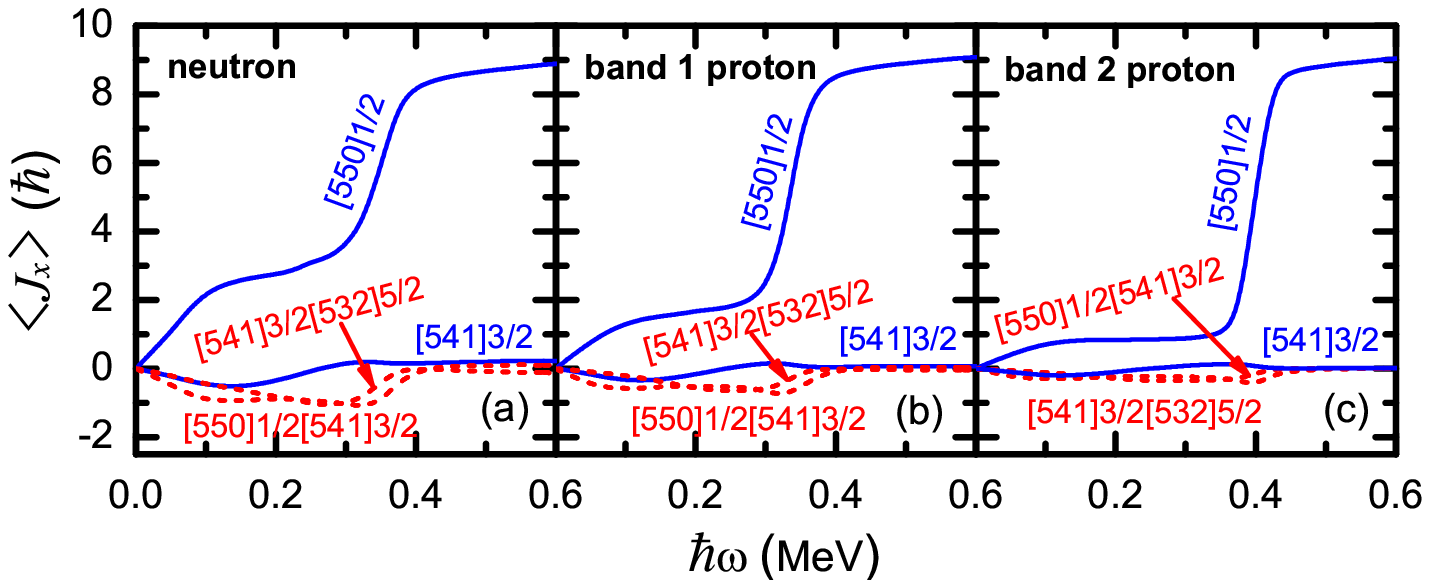}
\figcaption{\label{fig5:Jxorb}(Color online)
Contributions of each proton and neutron orbital in the $N=5$
major shell to the angular momentum alignments
$\langle J_x\rangle$ for the band 1 ($\pi 3/2^+[422], \alpha = -1/2$)
and band 2 ($\pi 1/2^+[420], \alpha = 1/2$) in $^{113}$Cs.
The diagonal (off-diagonal) part $j_x(\mu)$
[$j_x(\mu\nu)$] in Eq.~(\protect\ref{eq:jx}) is denoted by blue
solid (red dotted) lines.%
}
\end{center}
\begin{multicols}{2}

\section{Summary}
\label{sec:summ}

The recently observed two high-spin rotational bands in the proton emitter
$^{113}$Cs are investigated using the cranked shell model
with pairing correlations treated by a particle-number conserving method,
in which the Pauli blocking effects are taken into account exactly.
The effective pairing interaction strengths are determined by the experimental
odd-even differences in nuclear binding energies.
For each rotational band, the experimental MOI's and alignments can be reproduced
very well by the PNC-CSM calculations,
which in turn strongly support the configuration assignments for
band 1 ($\pi 3/2^+[422], \alpha = -1/2$)
and band 2 ($\pi 1/2^+[420], \alpha = 1/2$).
By analyzing the occupation probability $n_\mu$ of each cranked
Nilsson orbitals near the Fermi surface and the contribution of
each orbital to the angular momentum alignments,
the mechanism of the backbending in band 1 and band 2 can be understood clearly.

\end{multicols}

\vspace{-1mm}
\centerline{\rule{80mm}{0.1pt}}
\vspace{2mm}

\begin{multicols}{2}

\end{multicols}

\clearpage
\end{document}